\documentstyle[psfig]{mn2e}
\def\kms{km~s$^{-1}$}
\def\cm2{cm$^{-2}$}
\def\lya{Ly$\alpha$}
\def\nhi{$N$(H~I)}
\def\hkpc{$h_{70}^{-1}$ kpc}

\title[An excess of DLAs near QSOs.]{An excess of damped Lyman $\alpha$ 
galaxies near QSOs.}
\author[D. M. Russell, S. L. Ellison \& C. R. Benn]
{David M. Russell$^{1,2}$\thanks{Email: davidr@phys.soton.ac.uk} 
Sara L. Ellison$^3$ and Chris R. Benn$^1$
\\
$^1$Isaac Newton Group, Apartado 321, E-38700 Santa Cruz de La Palma, Spain\\
$^2$School of Physics \& Astronomy, University of Southampton, Highfield, 
Southampton, SO17 1BJ, United Kingdom\\
$^3$Dept. Physics \& Astronomy, University of Victoria, 3800 Finnerty Rd, 
Victoria, BC, V8P 1A1, Canada\\
}

\begin{document}
\maketitle

\begin{abstract}
We present a sample of 33 damped Lyman alpha systems (DLAs) discovered 
in the Sloan Digital Sky Survey (SDSS) whose absorption redshifts 
($z_{\rm abs}$)
are within 6000 \kms\ of the QSO's systemic redshift ($z_{\rm sys}$).
Our sample is based on 731  $2.5 < z_{\rm sys} <$ 4.5
non-broad-absorption-line (non-BAL) QSOs from Data Release 3 (DR3)
of the SDSS.  We estimate that our search is $\approx$100\% complete for 
absorbers with \nhi\ $\geq 2 \times 10^{20}$ \cm2.
The derived number density of DLAs per unit
redshift, $n(z)$, within $\Delta v < 6000$ \kms\ is higher (3.5 $\sigma$
significance) by almost a factor of 2 than that of intervening absorbers
observed in the SDSS DR3, i.e. there is evidence for an overdensity of
galaxies near the QSOs.  This provides a physical motivation for excluding DLAs
at small velocity separations in surveys of intervening `field'
DLAs.  In addition, we find that the overdensity
of proximate DLAs is independent of the radio-loudness of the QSO, 
consistent with the environments of radio-loud and radio-quiet QSOs being
similar.

\end{abstract}


\section{Introduction}
\lya\ absorption lines in the spectra of high redshift QSOs
trace the distribution of neutral hydrogen along the lines
of sight to the background sources.  Studying the properties of
these intervening absorption systems can provide insights into a 
range of cosmic
structures, ranging from the intergalactic medium to protogalaxies.
The absorbers with the highest column densities give rise to lines with 
characteristic Lorentzian profiles. Specifically, damped \lya\ 
(DLA) systems, defined as having \nhi\ $\ge 2 \times 10^{20}$ \cm2, 
dominate the mass density of HI in the universe and are believed to
be associated with the progenitors of present day galaxies.
Studies of DLAs have furnished a 
wealth of information about the density and chemical enrichment of
galaxy-scale clouds of neutral hydrogen
at early epochs (see for example the reviews by Pettini 2004;
Wolfe, Gawiser \& Prochaska 2005).  

Nearly all existing studies of DLAs have been confined to absorbers lying 
beyond a certain velocity from the systemic
redshift of the QSO ($z_{\rm sys}$).  The velocity cut imposed is somewhat
subjective and survey
dependent, but is usually 3000 -- 6000 \kms\ blueward of $z_{\rm sys}$.
The motivation for excluding  proximate DLAs ($\Delta v < 6000$ \kms, PDLAs)
is two-fold.  First, studies of the \lya\ forest have revealed the existence of
a `proximity effect', whereby the high UV flux from the QSO causes 
additional ionisation of the diffuse HI clouds within a few Mpc of the 
QSO (e.g. Bajtlik, Duncan \& Ostriker 1988; Lu, Wolfe \& Turnshek 1991).  
Even high column density systems whose interstellar media (ISM)
are usually considered to be self-shielded can be affected by proximity
to a powerful ionising source such as a QSO.  For example, it has
been noted that PDLAs seem to preferentially exhibit \lya\ emission 
superimposed on the \lya\ absorption trough (e.g. Ellison et al. 2002 and
references therein).  More recently, Adelberger et al. (2005)
have detected \lya\ fluorescence in a DLA separated by 380 (proper) \hkpc\  
from a $z = 2.84$ QSO.   Excluding absorbers which may
be affected by the QSO's ionising radiation simplifies the calculations
of its total hydrogen content and chemical abundances.  Second,
studies concerned with the statistical
properties of \textit{intervening} galaxies could be affected by the 
inclusion of material associated with the QSO.  Intrinsic absorption
could, for example, be associated with BAL-like outflows or with the 
QSO host galaxy.  However, the one $z_{\rm abs} \sim z_{\rm sys}$
PDLA with measured metal abundances is known to have a metallicity 
$\sim$ 15\% of the solar value (Meyer, Welty \& York 1989; Lu et al. 1996).
This value is typical of intervening DLAs but inconsistent with the
typically solar or super-solar metallicities of BAL-like outflows
and intrinsic systems (Barlow \& Sargent 1997; Petitjean, Rauch \& Carswell
1994; D'Odorico et al. 2004).  The PDLAs also lack the strong CIV or 
NV absorption and complex absorption profiles characteristic of 
BALs (see also M$\o$ller, Warren \& Fynbo 1998).
Taken together, this evidence implies that PDLAs are more likely to be 
related to gas associated with galaxies rather than AGN.
The differences in redshift between QSOs and PDLAs are typically too
large for the absorbing material to be part of the QSO host galaxy
suggesting that PDLAs may have an origin similar to 
the intervening systems, but are simply located closer to the QSO. 

In this paper, we consider whether there may be further justification
for excluding PDLAs from surveys by investigating whether they are 
drawn from the same `random' population as the intervening absorbers.  
It has been known for more than 30 years that QSOs are often associated 
with galaxy evolution (e.g. Bahcall, Schmidt \& Gunn 1969) so it
is plausible that PDLAs may represent a subset of early galaxies
in a distinct environment.  For example, it is
well-established that overdensities of galaxies are
often seen in the vicinity of QSOs (e.g. Yee \& Green 1987;
Ellingson, Green \& Yee 1991; 
Yamada et al. 1997; Hall \& Green 1998; Hall, Green \& Cohen 1998; 
Sanchez \& Gonzalez-Serrano 1999; McLure \& Dunlop 2001; Wold et al. 2001; 
Finn, Impey \& Hooper 2001; Clements 2000; Sanchez \& Gonzalez-Serrano 2002).  
Our objective is therefore to determine whether DLAs located at velocities
$<$ 6000 \kms\ from the background QSO
may similarly inhabit distinct environments compared with
intervening `field' absorbers.
As a first approach, we compare the number density of 
PDLAs to that of traditionally selected $z_{\rm abs} << z_{sys}$ DLAs.  
A number density of PDLAs consistent with that of intervening DLAs 
would imply no distinction in their respective environments.  
Conversely, an excess of DLAs near the QSO's systemic redshift
could be interpreted as probing galaxy overdensities analogous 
to those found by imaging surveys.  Overdensities of PDLAs would
provide an empirically motivated cut-off for intervening DLA
searches whose goal is to study an unbiased sample of absorbers.

Since the majority of DLA surveys exclude PDLAs, constructing a
statistical sample from the literature is challenging.  A handful
of PDLAs have been reported (e.g. M$\o$ller \& Warren 1993;
Pettini et al. 1995; M$\o$ller et al. 1998; Ellison et al. 
2002) but they have not been generally included in published lists of DLAs.
Recently, Ellison et al. (2002) searched for PDLAs in the Complete
Optical and Radio Absorption Line System (CORALS)  sample 
(see Ellison et al. 2001 for full sample definition).
Combining their sample with PDLAs discovered in previous 
surveys of radio loud and radio quiet QSOs (RLQs and RQQs respectively)
Ellison et al. (2002) found 4 PDLAs within $\Delta$v 
$<$ 3000 \kms\ of 96 radio-loud QSOs, and just 1 within $\Delta$v $<$ 
3000 \kms\ of 49 radio-quiet QSOs (P\'{e}roux et al. 2001).  
Given the redshift path ($\Delta z$)
covered by the RQQ sample, the detection of one PDLA is consistent
with the known number density per unit redshift ($n(z)$)
of intervening absorbers.
However, the detection of four PDLAs associated with RLQs
implies that the number density of PDLAs is $>$ 4 times that of the 
intervening population.  This apparent overdensity around RLQs at $z>2$ 
is significant only at the 1.5 $\sigma$ level due to the relatively 
small number of QSOs, and Ellison et al. (2002)
noted that better statistics are required to confirm the overdensity.

With the advent of the Sloan Digital Sky Survey (SDSS), the spectra of 
large numbers of QSOs ($\sim$ 10$^5$, eventually) are becoming available.
Here we search for PDLAs in the spectra of 
731 QSOs with apparent magnitude $i \sol 20.5$ from the SDSS 
Data Release 3 (DR3, Abazajian et al. 2005), in order
to answer the following two questions:

\begin{itemize}
\item Is the density of DLAs in the 
vicinity of QSOs significantly higher than that along the line of sight?
\item Do the environments of radio-quiet and radio-loud QSOs differ?
\end{itemize}

This work improves on that of Ellison et al. (2002) in several
ways.  First, our sample is an order of magnitude larger than the
CORALS survey.  With these improved statistics we can investigate
the number density in a range of velocity bins from $z_{\rm sys}$.
Since the data quality of the SDSS spectra is somewhat heterogeneous,
we have carried out completeness tests for our sample; these are 
described in \S3.  Finally, we
use a Monte Carlo simulation to assess the significance of the
measured overdensity of PDLAs relative to intervening systems.

\medskip

We have adopted the `consensus' cosmology model with 
$H_0=70$ \kms\ Mpc$^{-1}$, $\Omega_{\Lambda}$ = 0.7, $\Omega_{M}$ = 0.3.

\begin{table}
\centering
\caption{Metal-line rest-frame wavelengths}
\begin{tabular}{lc}
\hline
Line&Wavelength (\AA)\\
\hline
O I    &1302.17\\
Si IV  &1393.76\\
Si IV  &1402.77\\
Si II  &1526.71\\
C IV   &1548.20\\
C IV   &1550.78\\
Fe II  &1608.45\\
Al II  &1670.79\\
Al III &1854.72\\
\hline
\end{tabular}
\normalsize
\end{table}

\small
\begin{table*}
\caption{DLAs within 6000 \kms\ of the redshift of SDSS DR3 QSOs}
\begin{tabular}{rrrrrccrrl}
\hline\hline
\multicolumn{1}{c}{RA}&\multicolumn{1}{c}{Dec}&\multicolumn{1}{c}{$i$}&\multicolumn{1}{c}{z$_{\rm sys}$}&\multicolumn{1}{c}{S:N}&$S_{1.4GHz}$ mJy&\nhi&\multicolumn{1}{c}{z$_{\rm abs}$}&\multicolumn{1}{c}{$\Delta v$}&\multicolumn{1}{c}{Metal lines}\\
\multicolumn{1}{c}{J2000}&\multicolumn{1}{c}{J2000}&&&&(and log$_{10} R$)&$10^{20}$ \cm2&&\multicolumn{1}{c}{\kms}&\\
\multicolumn{1}{c}{(1)}&\multicolumn{1}{c}{(2)}&\multicolumn{1}{c}{(3)}&\multicolumn{1}{c}{(4)}&\multicolumn{1}{c}{(5)}&(6)&(7)&\multicolumn{1}{c}{(8)}&\multicolumn{1}{c}{(9)}&\multicolumn{1}{c}{(10)}\\
\hline \rule[0mm]{0mm}{4mm}
01 40 49.18&$-$08 39 42.5&17.7&3.713& 7.5&    -&4.5 &3.696&1085&OI SiII [BAL]                           \\ \rule[0mm]{0mm}{4mm}
07 39 38.85&+30 59 51.2&20.4&3.399& 9.2& 6.69 (2.9)&2.0 &3.353&3160&SiII CIV                 \\ \rule[0mm]{0mm}{4mm}
08 05 53.02&+30 29 37.3&19.9&3.432& 9.9&    -&2.5 &3.429&200 &OI SiII CIV                  \\ \rule[0mm]{0mm}{4mm}
08 11 14.32&+39 36 33.2&20.1&3.092& 6.9&    -&7.0 &3.037&4060&                        \\ \rule[0mm]{0mm}{4mm}
08 26 38.59&+51 52 33.2&17.2&2.850& 5.9&    -&5.5 &2.833&1330&OI SiIV  SiII CIV  FeII AlII      \\ \rule[0mm]{0mm}{4mm}
08 44 51.72&+05 18 27.8&20.0&4.465& 7.2&    -&3.5 &4.376&4925&                        \\ \rule[0mm]{0mm}{4mm}
09 09 30.42&+07 00 50.7&20.3&3.273& 5.5&    -&3.0 &3.219&3815&                        \\ \rule[0mm]{0mm}{4mm}
09 40 35.93&+50 03 08.7&20.1&3.567& 5.8&    -&2.5 &3.500&4435&                        \\ \rule[0mm]{0mm}{4mm}
10 11 22.59&+47 00 42.2&19.0&2.928& 9.1&29.56 (2.5)&2.5 &2.909&1490&OI SiII           \\ \rule[0mm]{0mm}{4mm}
10 26 19.09&+61 36 28.9&18.6&3.849& 5.3&    -&3.0 &3.785&3985&OI SiIV  SiII CIV  AlII       \\ \rule[0mm]{0mm}{4mm}
10 45 01.44&+50 40 45.9&20.0&3.998& 5.1&    -&6.0 &3.949&2955&                        \\ \rule[0mm]{0mm}{4mm}
10 48 20.93&+50 32 54.2&20.2&3.889&12.0&    -&15.0&3.815&4545&OI SiIV  CIV              \\ \rule[0mm]{0mm}{4mm}
11 12 24.18&+00 46 30.3&19.7&4.035&14.0&    -&19.0&3.958&4620&OI AlII                     \\ \rule[0mm]{0mm}{4mm}
11 26 48.62&+05 56 28.1&19.6&3.191&17.1&    -&7.0 &3.147&3165&                        \\ \rule[0mm]{0mm}{4mm}
12 27 23.65&+01 48 06.0&20.4&2.881&10.1&    -&2.0 &2.876&425 &                        \\ \rule[0mm]{0mm}{4mm}
12 42 04.27&+62 57 12.1&19.6&3.321& 5.7& 0.88 (1.4)&2.0 &3.276&3140&SiIV SiII CIV  FeII AlII         \\ \rule[0mm]{0mm}{4mm}
12 48 30.64&+49 14 00.2&20.4&3.079& 7.7&    -&4.0 &3.032&3460&                        \\ \rule[0mm]{0mm}{4mm}
12 57 59.22&$-$01 11 30.2&18.7&4.112& 8.1&    -&2.0 &4.022&5330&OI SiIV  CIV              \\ \rule[0mm]{0mm}{4mm}
13 25 21.27&+52 45 13.1&19.8&3.953& 6.3&    -&2.5 &3.861&5625&OI                      \\ \rule[0mm]{0mm}{4mm}
13 41 00.13&+58 07 24.2&19.4&3.500& 7.3&    -&7.0 &3.417&5585&all those listed in Table 1                   \\ \rule[0mm]{0mm}{4mm}
13 46 37.94&+56 49 15.6&19.7&3.463& 6.7&    -&3.0 &3.430&2225&OI                      \\ \rule[0mm]{0mm}{4mm}
14 00 29.01&+41 12 43.4&19.0&2.540& 5.9&    -&45.0&2.537&255 &OI SiII CIV  FeII AlII           \\ \rule[0mm]{0mm}{4mm}
14 20 41.96&+42 22 57.0&19.6&3.378& 8.2&    -&2.0 &3.336&2925&OI                      \\ \rule[0mm]{0mm}{4mm}
14 51 20.65&+39 13 50.4&20.4&3.376& 8.3&    -&2.0 &3.307&4770&SiII CIV                 \\ \rule[0mm]{0mm}{4mm}
15 12 54.37&$-$00 56 36.6&20.2&4.457& 5.6&    -&8.0 &4.400&3150&OI SiIV  SiII CIV  FeII       \\ \rule[0mm]{0mm}{4mm}
15 52 48.01&+56 03 28.9&20.0&3.604&12.4&    -&2.0 &3.556&3145&                        \\ \rule[0mm]{0mm}{4mm}
16 08 13.86&+37 47 27.3&20.3&3.996& 8.5&    -&2.0 &3.952&2655&OI SiIV                  \\ \rule[0mm]{0mm}{4mm}
16 11 19.56&+44 11 44.0&20.0&4.024& 5.1&    -&5.0 &3.982&2520&SiIV                     \\ \rule[0mm]{0mm}{4mm}
17 04 21.85&+62 47 41.5&20.1&2.980& 6.0&    -&6.0 &2.953&2040&OI SiIV                     \\ \rule[0mm]{0mm}{4mm}
17 18 00.20&+62 13 25.6&19.8&3.670& 7.8&    -&3.5 &3.618&3360&OI SiII FeII AlII               \\ \rule[0mm]{0mm}{4mm}
17 20 07.20&+60 28 23.8&20.4&4.425& 8.6& 5.06 (2.2)&5.0 &4.326&5525&                        \\ \rule[0mm]{0mm}{4mm}
21 00 25.03&$-$06 41 46.0&18.1&3.138& 9.0&    -&9.0 &3.092&3350&all those listed in Table 1                \\ \rule[0mm]{0mm}{4mm}
21 22 07.36&$-$00 14 45.7&19.0&4.072& 5.2&    -&2.5 &4.001&4230&OI CIV                   \\
\hline
\end{tabular}

\normalsize
The columns give: (1-4) SDSS RA, Dec, i mag and emission-line redshift;
(5) S:N in the continuum redward of the \lya\ emission;
(6) FIRST 1.4-GHz radio flux density (and log$_{10} R$ where R is the radio-loudness parameter (see \S2);
log$_{10} R >$ 1.0 for a RLQ);
(7) fitted H I column density of the DLA;
(8) DLA redshift;
(9) DLA velocity, relative to $z_{\rm sys}$;
(10) metal lines associated with the DLA

\end{table*}

\normalsize
\section{The sample}
SDSS DR3 covers $\sim$5300 deg$^2$, mainly in the northern sky, and
includes spectra of 5$\times10^{4}$ QSOs with $i$ $\la$ 20.5. 
From these, we selected 1885 2.5 $< z_{\rm sys} <$ 4.5 QSOs.
The lower redshift limit was selected so that at least 250 \AA\ of
the \lya\ forest was covered by the SDSS spectrum (whose blue limit 
is $\approx$ 4000 \AA).  The upper redshift limit 
was imposed to avoid the difficulty of identifying DLAs unambiguously 
in the much noisier and densely populated \lya\ absorption forests of QSOs
at higher redshift.  The 1885 spectra were inspected individually, and 413 
objects showing broad absorption lines just blueward of the CIV emission 
line were rejected.  
These are probable broad-absorption-line (BAL) QSOs, and a search for 
proximate DLAs could be confused by HI absorption features associated with 
the BAL clouds\footnote{This selection left within the sample one object 
meeting the formal
criteria for a BAL (0140$-$0839) but with the absorption detached $\sim$ 27000 
\kms\ blueward of the CIV emission, i.e. not contaminating the part of 
the spectrum searched for PDLAs.}. BAL QSOs were rejected
on the basis of visual inspection, and all three authors independently checked the
classifications.

We rejected 665 spectra with continuum
signal-to-noise ratio per pixel (S:N) less than 5.0 
in the continuum redward of the \lya\ emission, based on the
S:N estimates (in g, r and i bands) included in the headers of the 
SDSS spectra.  We also measured the S:N directly 
from the spectrum in a sample of 50 QSOs and found that 90\% of our 
measured S:N agree with the quoted S:N within 0.2.

\begin{figure*}
\centering
\psfig{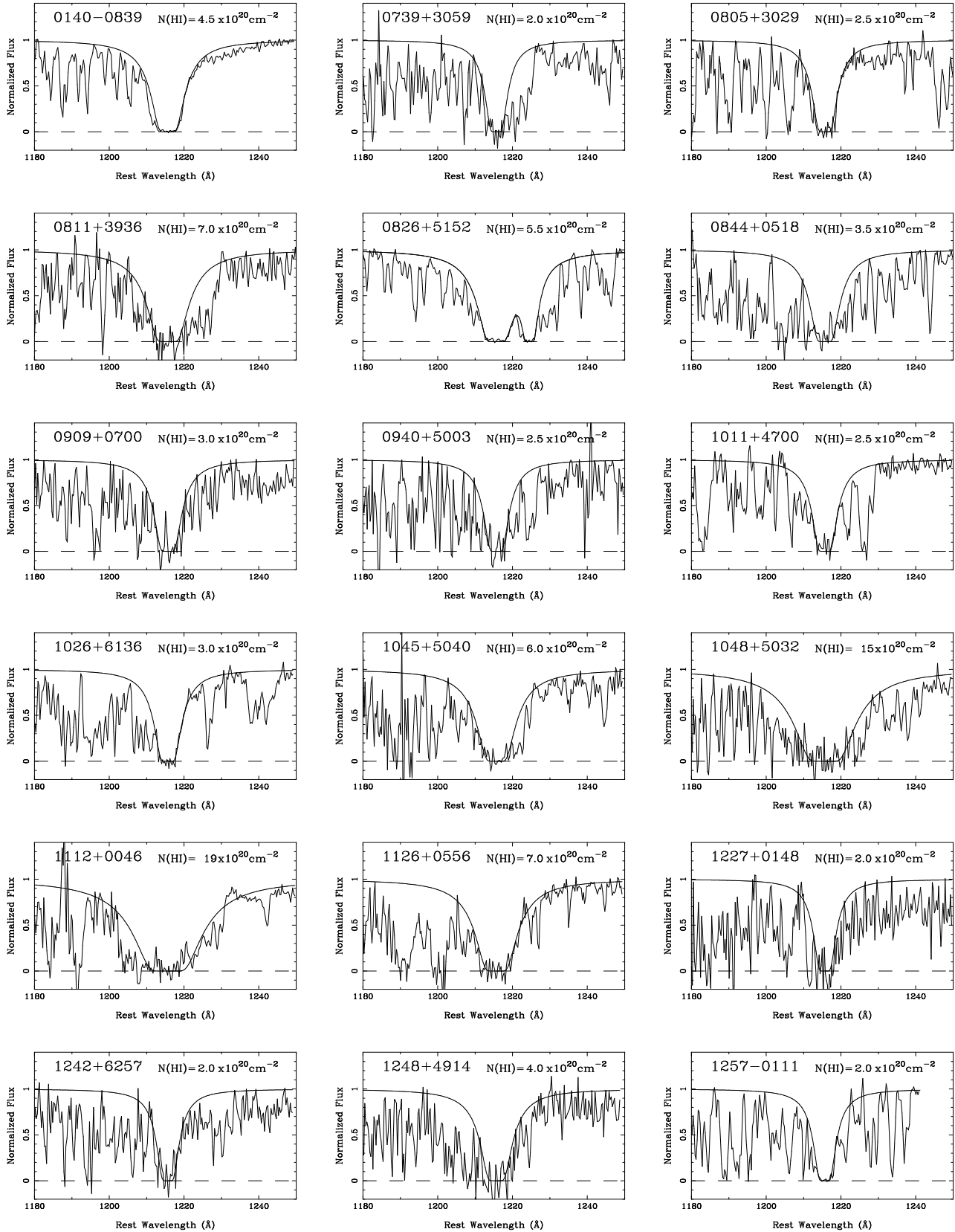}
\caption{\label{dla_fit} Theoretical damped \lya\ profile fits to all 33 
PDLAs.  The fitted \nhi\ is indicated at the top right of each plot.  Table 2 gives the derived $z_{\rm abs}$ for each absorption system.
For QSO 0826$+$5152, the fit is for the two absorbers, and the wavelength scale is in the rest frame of the blueward feature, which is the PDLA in this system.  The HI column density of the redward absorber is 1.3 $\times$ 10$^{20}$ \cm2.}
\end{figure*}

\begin{figure*}
\centering
\psfig{file=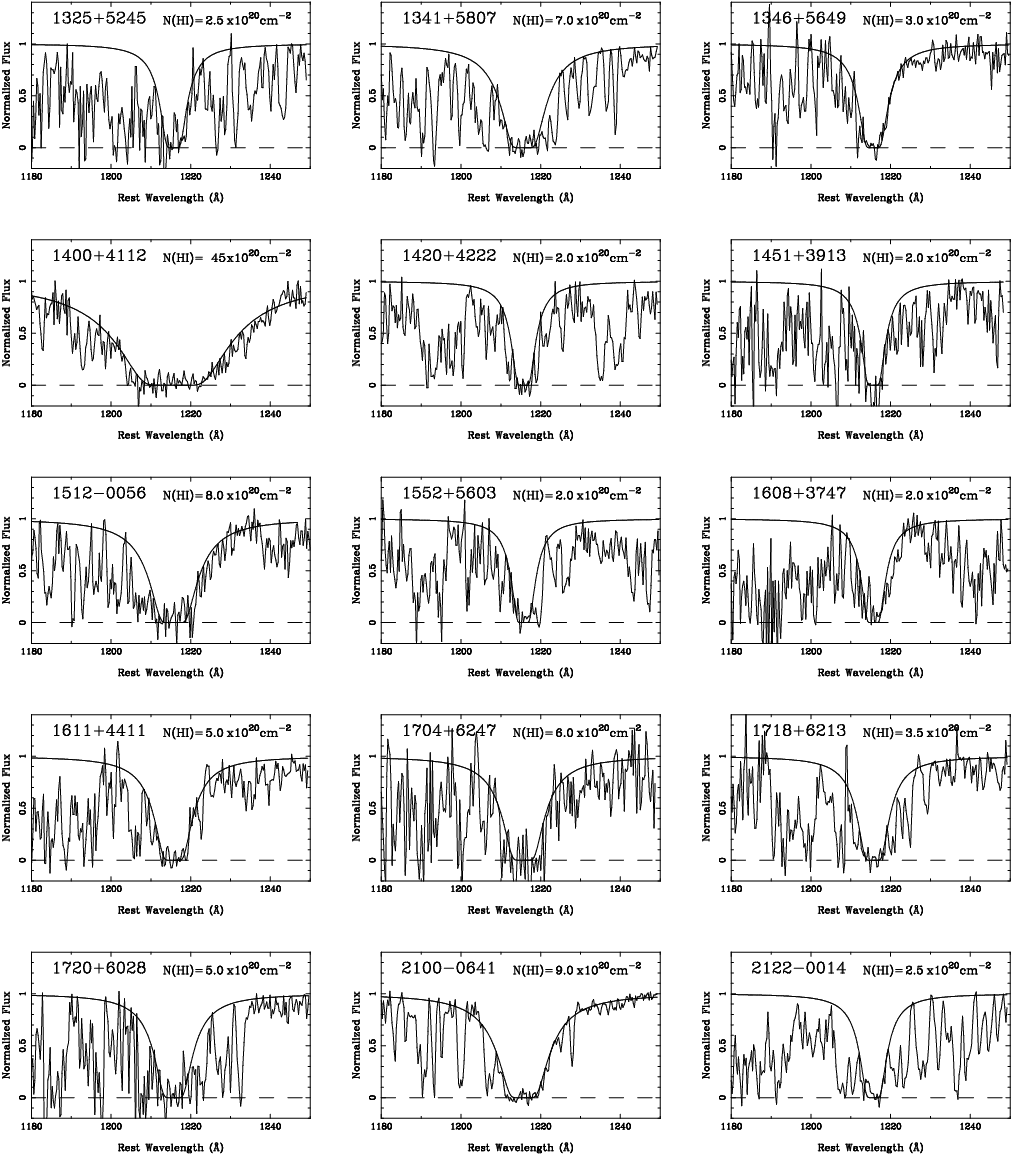,width=17cm,height=18.2cm,angle=0}
\contcaption{ }
\end{figure*}

Finally, to ensure availability of radio flux densities (or limits) for the
QSOs, we rejected the 76 QSOs not falling within areas covered by the
FIRST survey (Faint Images of the Radio Sky at Twenty cm, 
$S_{1.4GHz} \sog$ 1 mJy; Becker, White \& Helfand 1995).  This left a final
sample of 731 QSOs satisfying the following criteria: 
2.5 $< z_{\rm sys} <$ 4.5; $i \sol$ 20.5 (the SDSS limit); 
continuum S:N $>$ 5.0 redward of 
the \lya\ emission; non-BAL; and within the area 
covered by FIRST.

Radio counterparts were sought in the FIRST catalogue and sources
within 2 arcsec of the optical position were assumed 
to be true radio identifications.  The radio images of other sources 
lying within 30 arcsec of the QSO (9 in all) were inspected individually.  
None of these lie $\leq$ 5 arcsec from a line joining the two radio
counterparts, and so the 9 are assumed to be unrelated sources.
Based on the FIRST source density, we estimate the expected 
number of chance radio--optical coincidences to be 0.06.
Following Stocke et al. (1992), we take as our definition of radio loud: 
log$_{10} R >$ 1.0, where $R$ (radio-loudness parameter) is the 
ratio of K-corrected radio and optical flux densities: 
$F_\nu$(5GHz)/$F_\nu$(2500\AA).  The 
5-GHz radio flux density was obtained from the FIRST 1.4-GHz flux density,
assuming a mean radio 
spectral index $\alpha$ = $-0.3$ ($S_\nu \propto \nu^{\alpha}$).
Of the 731 QSOs in the sample, 94 are radio-loud and 637 are 
radio-quiet.

\section{Identification of PDLAs}

We searched each of the 731 SDSS spectra 
for absorption features satisfying the following criteria:

\begin{itemize}
\item velocity $<$ 6000 \kms\ blueward of the \lya\ emission;
\item saturated absorption trough (i.e. minimum consistent with zero 
residual flux);
\item \nhi\ $\geq 2.0\times10^{20}$ \cm2\ 
(corresponding to rest-frame equivalent width $\sog$ 10\AA).
\end{itemize}

The HI column densities, \nhi, of the candidates were measured
by fitting theoretical \lya\ absorption profiles to the
spectra 
using the DIPSO package (Howarth et al. 2003), iterating 
in \nhi\ and redshift until a good fit was found (usually after
two iterations).  
We also searched for the strong metal absorption lines commonly associated 
with DLAs (Table 1) to help constrain the redshift.
The search yielded 33 proximate DLAs (Table 2).
Spectra of the PDLAs, as well as the \nhi\ fits, are shown in Fig. 
\ref{dla_fit}. 

\begin{figure}
\centering
\psfig{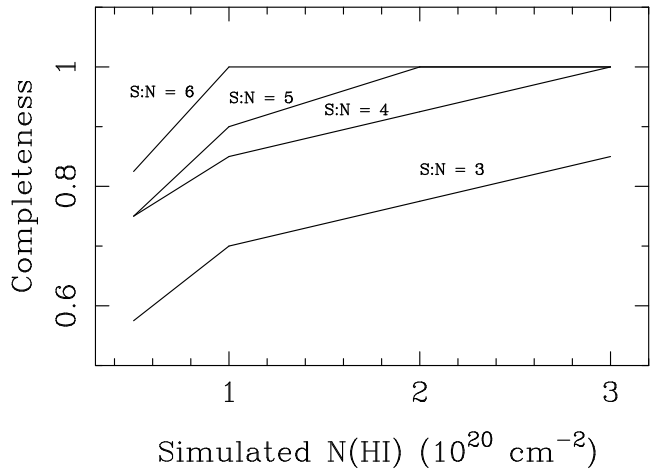}
\caption{\label{complete_test} The fraction of simulated DLAs that were identified,
as a function of the S:N on the quasar spectrum and
the HI column density.
For S:N $\geq$ 5 and \nhi\ $\geq$ 2.0$\times$10$^{20}$ \cm2,
the completeness of the search is $\approx$ 100\%.
}
\end{figure}

A concern when working with moderate--to--low S:N spectra
is the level of completeness in the absorber sample, either
due to blending or
underestimating the column densities near to the $2 \times 10^{20}$ \cm2\ 
limit.  To quantify the completeness of our sample, we added simulated 
DLAs to real spectra of QSOs (blueward of \lya) and attempted to
recover them, adopting the same criteria as we used for the search for real
DLAs.  We simulated a range of column densities \nhi\ = 0.5, 1.0, 
2.0 and 3.0 $\times10^{20}$ \cm2, and S:N per pixel = 3, 4, 5 and 6 
(measured redward of \lya).  For each combination of S:N and \nhi\, 
we created 40 simulated spectra, i.e. 640 spectra in all.
The simulations and searches were made by different members of the team;
the results are shown in Fig. \ref{complete_test}. 
All of the simulated DLAs with S:N
$\geq$ 5.0 and $N(H I)$ $\geq$ 2.0 $\times10^{20}$ \cm2\ were 
identified and recovered, indicating that our search for DLAs with
\nhi\ $\geq$ 2.0
$\times10^{20}$ \cm2\ and S:N $\geq$ 5 is likely to be 100\% complete.  
For S:N = 4, the completeness would be 93\%.

Uncertainties in the \nhi\ determinations come from both statistical
and systematic (e.g. continuum fitting) errors.  
Prochaska, Herbert-Fort \& Wolfe (2005) independently discovered, 
and measured the \nhi\ of, 17 of the PDLAs in Table 2. 16 of these 
measurements of \nhi\ agree with our values, within 0.2
dex.  The column densities are not systematically over- or under-estimated
by either group, so we consider 0.2 dex a realistic fiducial error
which accounts for both statistical and systematic uncertainties.

Notes on individual PDLAs:
\newline\newline
\textbf{SDSS J014049.18$-$083942.5}
This velocity-detached BAL QSO has a PDLA with strong O I and Si II 
absorption lines.  These metal lines confirm the redshift derived from a
good fit to the Voigt profile.
\newline\newline
\textbf{SDSS J080553.02+302937.3}
The relatively high signal-to-noise ratio (S:N = 9.9 in the continuum redward of
\lya) produces an excellent profile fit, with little blending from
neighbouring absorbers.  Strong O I and C IV absorption defines the DLA
redshift at just 200 \kms\ blueward of the \lya\ emission line. With such a
small relative velocity, this absorber could be associated with the QSO host galaxy. This is an
interesting source to follow up with high resolution spectroscopy.
\newline\newline
\textbf{SDSS J082638.59+515233.2}
This relatively bright QSO ($i =$ 17.3) 
possesses a PDLA with a well-defined fit, particularly in the trough,
with \nhi\ =  5.5 $\times 10^{20}$ \cm2.
Many strong metal lines are associated with the absorber; all those listed
in Table 1 except the Al III line. 
A sub-DLA, \nhi\ = 1.3 $\times$ 10$^{20}$ \cm2,
is evident just to the red of the main DLA, velocity
$\sim -850$ \kms\ (i.e. redward) of the QSO redshift.
Three metal lines are detected in association with the sub-DLA:
the blue component of the
C IV doublet, and both components of the Si IV doublet.
The \lya\ emission line appears to be suppressed by these absorbers.
A handful of multiple DLAs have been reported at intervening redshifts
(e.g. Ellison \& Lopez 2001; Lopez \& Ellison 2003; Prochaska \& Herbert-Fort 
2004; Turnshek et al. 2004). Statistically, these are
improbable events if the proximity of the systems is purely by
chance, and they may trace large-scale
structures (e.g. Ellison \& Lopez 2001; Turnshek et al. 2004). The
double DLA found towards this QSO has a separation $\Delta$v $\sim$ 2180 \kms.
\newline\newline
\textbf{SDSS J101122.59+470042.2}
This RLQ (the brightest at radio wavelengths; $S_{1.4GHz}\sim30$ mJy)
possesses a PDLA with heavy blending in the wings of the profile. However a
good fit to the profile in the trough constrains the \nhi\ and redshift, the
latter of which is confirmed by a strong O I line.
\newline\newline
\textbf{SDSS J102619.09+613628.9}
The DLA is detected in all the metal lines listed in Table 1, except 
the Al III line, which falls outside the range of the spectrum. 
The absorber is well-fitted by the Voigt profile in the trough and
on the red wing, with narrow absorbers contaminating the fit to the blue wing.
\newline\newline
\textbf{SDSS J104820.93+503254.2}
The redshift of this high \nhi\ absorber cannot be accurately measured from the
profile fit due to narrow blended absorbers, however 5 associated metal
lines constrain the redshift in this high-S:N (12.0) spectrum.
\newline\newline
\textbf{SDSS J111224.18+004630.3}
One of the highest column density 
PDLAs known (\nhi\ = 19$\times 10^{20}$ \cm2), the
redshift is well-constrained by the detection of O I and Al II lines. Many deep
absorbers confuse the fit on the blue side of the trough, although the profile
fits well on the red side and on both wings.  However, this system exhibits the
largest discrepancy with the \nhi\ determined by Prochaska et al. (2005):
3.5$\times 10^{20}$ \cm2.  This is a good example of the significant errors that
can arise when estimating \nhi\ for DLAs which are heavily blended.
\newline\newline
\textbf{SDSS J134100.13+580724.2}
A collection of very prominent metal lines (all those listed in Table 1) define the
redshift of this PDLA. Despite contamination from deep neighbouring absorbers
in the \lya\ forest, the \nhi\ is well-constrained by the fit in the trough
and on the red wing.
\newline\newline
\textbf{SDSS J140029.01+411243.4}
This PDLA not only possesses the largest column density reported in this
paper (well fitted by a Voigt profile with 45$\times 10^{20}$
\cm2), but is also just $\sim$ 250 \kms\ blueward of the \lya\ emission. The
small velocity relative to the QSO suggests that the absorber may be associated
with the host galaxy.  Measurements of the absorber's
metallicity may reveal its nature; many metal lines are seen in the SDSS
spectrum, but the high equivalent widths indicate that
they are all saturated and therefore not useful for abundance determinations.
\newline\newline
\textbf{SDSS J155248.01+560328.9}
The estimated column density of $2.0\times 10^{20}$ \cm2\ produces an
acceptable fit, however a deep but narrow absorption line blends with the
blue wing.  Strong C IV absorption (1548 \AA, 1550 \AA) is detected at
the redshift of this second feature, but no metal lines are detected in
association with the PDLA.
\newline\newline
\textbf{SDSS J171800.20+621325.6}
Two narrow but deep absorbers blend with the wings of this PDLA, but the redshift and column density are well-defined, by the fit in the trough and by four associated metal lines; O I, Si II, Fe II and Al II.
\newline\newline
\textbf{SDSS J210025.03$-$064146.0}
This absorber has a high column density (\nhi\ = 9.0 $\times 10^{20}$ \cm2)
and all the metal lines listed in Table 1 are detected.

\small
\begin{table*}
\centering
\caption{Number density of DLAs per unit redshift, $n(z)$, for 
different samples and velocity ranges}
\begin{tabular}{clccccccccc}
\hline\hline
&Sample&QSO&No. of&$\Delta v$ limit&$N_{PDLA}$&$\Delta z_v$&$n(z)_{PDLA}$&$\langle z_{\rm abs}\rangle$&$N_{DLA}$&overdensity\\
&&Type&QSOs&(\kms)&&&&&expected&\\
&(1)&(2)&(3)&(4)&(5)&(6)&(7)&(8)&(9)&(10)\\
\hline \rule[0mm]{0mm}{4mm}
(1)&SDSS&RQQ&637&6000&29&55.02&0.53$^{+0.12}_{-0.10}$&3.521&16.0&1.8\\ \rule[0mm]{0mm}{4mm}
(2)&SDSS&RQQ&637&3000&11&27.64&0.40$^{+0.16}_{-0.12}$&3.361& 7.7&1.4\\
\hline \rule[0mm]{0mm}{4mm}
(3)&SDSS&RLQ&94&6000&4&7.74&0.52$^{+0.41}_{-0.25}$&3.466& 2.2&1.8\\ \rule[0mm]{0mm}{4mm}
(4)&SDSS+CORALS+FBQS&RLQ&190&3000&5&7.47&0.67$^{+0.45}_{-0.29}$&2.790& 1.8&2.8\\
\hline \rule[0mm]{0mm}{4mm}
(5)&SDSS&All&731&6000&33&62.76&0.53$^{+0.11}_{-0.09}$&3.515&18.2&1.8\\ \rule[0mm]{0mm}{4mm}
(6)&All Above Surveys&All&827&3000&16&35.11&0.46$^{+0.14}_{-0.11}$&3.182& 9.5&1.7\\
\hline \rule[0mm]{0mm}{4mm}
(7)&SDSS&RQQ&40& none &9&40.52&0.22$^{+0.10}_{-0.07}$&3.08&10.5&0.8\\
\hline
\end{tabular}

\normalsize
The columns give: 
(1) PDLA sample = SDSS (this paper), the CORALS Survey (Ellison et al. 2002), 
FBQS (FIRST Bright Quasar Survey, White et al. 2000);
(2) QSO type, radio-quiet or radio-loud;
(3) number of QSOs in the sample;
(4) velocity limit;
(5) number of DLAs found within given $\Delta v$;
(6) total redshift path $\Delta z$ searched;
(7) number density $n(z)$ and statistical error;
(8) mean redshift of DLAs;
(9) number of intervening DLAs predicted for a redshift path $\Delta z_v$,
using the formula of Storrie-Lombardi \& Wolfe (2000);
(10) overdensity of PDLAs $N_{PDLA}$ / $N_{DLA}$. The statistical significance
of there being an overdensity is 3.0$\sigma$, 1.4$\sigma$, 1.4$\sigma$ and 3.5$\sigma$
in rows 1, 2, 3 and 5 respectively, as estimated from Monte Carlo
simulations (see \S4).
Row (7) gives the result of our search for intervening DLAs in 40 
SDSS spectra (a check of our search method).
\end{table*}

\normalsize
\section{THE DLA NUMBER DENSITY}
The number density of DLAs per unit redshift n(z) is obtained by
dividing the number of DLAs found by the total
redshift path sampled, $\Delta z$ and is quoted for a given
mean redshift:
\begin{eqnarray*}
  \Delta z=\sum\limits^N_{i=1} (z_{\rm max,i}-z_{\rm min,i})
\end{eqnarray*}

where $z_{\rm max,i}$ here is the QSO systematic redshift $z_{\rm sys,i}$,
$z_{\rm min,i}$ is the redshift corresponding to a given velocity limit, and
the summation is over the $N$ QSOs in the sample.  
For the samples of radio-quiet 
and radio-loud QSOs, $N=637$ and 94 respectively.  
The values of $\Delta v$, $z_{\rm min,i}$ and 
$z_{\rm sys,i}$ are related by:

\begin{eqnarray}
  r=\frac{1+z_{\rm sys,i}}{1+z_{\rm min,i}}\\
  \frac{\Delta v}{c}=\frac{r^2-1}{r^2+1}
\end{eqnarray}
\newline
Equation (1) can be rewritten:
\begin{eqnarray}
  z_{\rm sys,i}-z_{\rm min,i}=\left(1-\frac{1}{r}\right)(1+z_{\rm sys,i})
\end{eqnarray}
\newline
We measure $n(z)$ for the velocity ranges (1) $\Delta$v$< 6000$ \kms,
and (2) $\Delta$v$< 3000$ \kms\ (which is the most frequently used velocity
cut-off in DLA surveys).
For $\Delta v <$ 6000 \kms\ ($r$ = 1.0202),
equation 3 yields the total redshift intervals searched for 
the 637 radio-quiet and 94 radio-loud QSOs:
\begin{eqnarray*}
  \Delta z_{\rm rq}=\sum\limits^{637}_{i=1} (z_{\rm sys,i}-z_{\rm min,i})=55.02\\
  \Delta z_{\rm rl}=\sum\limits^{94}_{i=1} (z_{\rm sys,i}-z_{\rm min,i})=7.74
\end{eqnarray*}
\newline
We discovered 29 radio-quiet QSOs with PDLAs so the number density for radio-quiet QSOs is:
\begin{eqnarray*}
  n(z)_{\rm rq}=\frac{29}{55.02}=0.53^{+0.12}_{-0.10}
\end{eqnarray*}
 (using 1$\sigma$ confidence limits given by the Poissonian small number 
statistics; Gehrels 1986) and the number density for the radio-loud QSOs
(4 PDLAs discovered) is:
\begin{eqnarray*}
  n(z)_{\rm rl}=\frac{4}{7.74}=0.52^{+0.41}_{-0.25}
\end{eqnarray*}
Repeating the calculations for a search limit $v <$ 3000 \kms\ ($r$ =
1.0101), we find $\Delta z_{\rm rq}$ = 27.64, and $\Delta z_{\rm rl}$ = 3.89.  
We discovered 11 radio-quiet QSOs and 
1 radio-loud QSO with 
PDLAs within 3000 \kms\ (see Table 2), so the implied $n(z)$ are 
$n(z)_{\rm rq}=0.40^{+0.16}_{-0.12}$ and $n(z)_{\rm rl}=0.26^{+0.59}_{-0.21}$ per 
unit redshift.  
We compare in Table 3 our measured values of $n(z)$ for PDLAs
with those from previous surveys, and with those predicted for intervening DLAs by
Storrie-Lombardi \& Wolfe's (2000)
empirical formula $n(z)=0.055(1+z)^{1.11}$.
The $n(z)$ of the complete DR3 sample 
of intervening DLAs is in good agreement with Storrie-Lombardi \& Wolfe over
the redshift range studied here (J. X. Prochaska, private communication).

\begin{figure}
\centering
\psfig{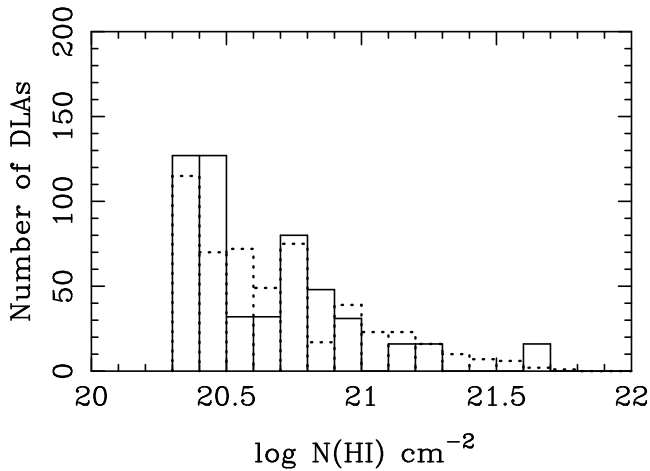}
\caption{\label{nhi_hist} Distribution of \nhi\ 
column densities of the PDLAs discovered here (solid line) compared with
that of the intervening DLAs from SDSS DR3 (Prochaska et al. 2005;
dashed line). The y-axis refers to the number of DLAs in the sample of
Prochaska et al. (total 525). To facilitate comparison, our sample of PDLAs
has been normalised such that it has the same total area under the histogram.}
\end{figure}

As a check of our method of measuring $n(z)$, 
we searched for intervening DLAs (with any velocity) in 40 
randomly-selected SDSS QSOs with a distribution of redshifts similar to 
that in Table 2.  
We found nine (row 7 of Table 3), 
with a distribution of column densities consistent with that in Fig. 
\ref{nhi_hist}.  
The implied $n(z)$ is 0.22$^{+0.10}_{-0.07}$, in good agreement with the 
value of 0.26 predicted by Storrie-Lombardi \& Wolfe,
for $\langle z_{\rm abs}\rangle$=3.08.

To compare the distribution of PDLA \nhi\ column densities with that
reported for intervening DLAs in the SDSS DR3 (from Prochaska et al.
2005), we applied a Kolmogorov-Smirnov (K-S) test.
The K-S probability that the distributions of PDLA and
intervening DLA \nhi\ column densities differ is 93\% (the maximum
difference between the cumulative distributions is 0.23), slightly
less than a 2 $\sigma$ result.  However, K-S test results require at least
a 95\% probability before they are usually considered to provide evidence
against the null hypothesis.  It is therefore uncertain from
Fig. \ref{nhi_hist} and the K-S test results whether the 
distributions differ in a 
systematic way.  In \S5.2 we return
to the issue of \nhi\ differences between PDLAs and DLAs.

A possible source of error in the measured $n(z)$ (which could
also affect the K-S test) is the error in measuring \nhi,
which is particularly critical near the $2\times 10^{20}$ \cm2\ cut-off.  
If the number of \nhi\ values (Fig. \ref{nhi_hist}) 
rises as \nhi\ decreases, then
measurement error could promote more \nhi\ values above the limit of
$2.0\times 10^{20}$ \cm2\ than are pushed below it (a Malmquist bias; e.g. Rao \& Turnshek 2000).  
Using the \nhi\ column-density distribution given in Fig. 4 of 
Ellison et al. (2001), we estimate that the 
measured number of DLAs in the range 20.3 $\leq$ log $N(H I)$ $<$ 20.4 will exceed the true value
by only $\sim$  3\%, i.e. the error in $n(z)$ is negligible compared to statistical noise.
For the 17 DLAs in common with the Prochaska et al. (2005) work, we see that
not only are the column densities generally in good agreement (80\% agree within
$\pm$ 0.1 dex), but there is no systematic over- or under-estimate by either
group.  Measurement errors are therefore unlikely to dominate the K-S test.

Finally, we investigated the possibility that our measurements of $n(z)$
might depend on the redshift range used, because 
the \lya\ forest is usually noisier in high-redshift QSOs, 
making it
more difficult to fit profiles to candidate DLAs.  
We recalculated $n(z)$ for QSOs with $z_{\rm sys}\leq$ 3.5, removing from
the sample the 16 
highest redshift PDLAs with $\Delta z<$ 6000 \kms.
The 17 remaining low-redshift PDLAs with $\Delta z<$ 6000 \kms\
give $n(z)$ = 0.53$^{+0.16}_{-0.13}$ (with redshift range $\Delta 
z_v$=32.14), and the 6 low-redshift PDLAs with $\Delta z<$ 3000 \kms\ 
yield $n(z)$ = 0.37$^{+0.22}_{-0.15}$ ($\Delta z_v$=16.15). 
These values of $n(z)$ are consistent with $n(z)$ measured using all
the PDLAs, implying that the inclusion
of high redshift systems does not significantly bias our result.

\vspace{5mm}
We find (Table 3) that the number densities of PDLAs $n(z)_{PDLA}$ exceed the
expected values for intervening DLAs $n(z)_{DLA}$ given by Storrie-Lombardi 
\& Wolfe (2000).  To estimate the significance of these apparent overdensities
we need to consider the errors associated with the expected number density for
intervening systems, as well as the Poisson errors quoted for the
PDLAs.  For the intervening DLAs we determine the statistical error on
$n(z)$ using Monte Carlo simulations of our QSO catalogue. 
We start by determining 1 $\sigma$ confidence limits (Gehrels 1986)
associated with $n(z)_{DLA}$ as a function of redshift from Table 9 of
Storrie-Lombardi \& Wolfe (2000).  The results are shown in Table 4.  We then
took 731 QSOs with  a redshift distribution identical to our SDSS sample and
allocated a number density $n(z)_{rand}$ drawn from a Gaussian distribution,
$n(z)_{DLA} \pm \sigma$, for each QSO.  Using Poisson statistics, the
probability of finding  $m$ DLAs for a given QSO was calculated; 
$P(m)=e^{-N}N^m/m$!, where $N$ is the expected number of DLAs for a given QSO;
$N=N_{DLA}=n(z)_{rand}\Delta z_v$.  In this way, we determined the
number of intervening DLAs, $N_{DLA}$, expected at $\Delta v < 6000$ \kms.

\begin{table}
\centering
\caption{$n(z)_{DLA}$ from Storrie-Lombardi \& Wolfe (2000) and 1$\sigma$ confidence
limits determined by us (see \S4).}
\begin{tabular}{cccc}
\hline
redshift range&$\Delta z$&$N_{DLA}$&$n(z)_{DLA}$\\
\hline
2.5 - 3.0&76.9&15&0.20$^{+0.06}_{-0.05}$\\ \rule[0mm]{0mm}{4mm}
3.0 - 3.5&40.9&10&0.24$^{+0.11}_{-0.07}$\\ \rule[0mm]{0mm}{4mm}
3.5 - 4.7&33.8&12&0.36$^{+0.13}_{-0.11}$\\
\hline
\end{tabular}
\normalsize
\end{table}

\begin{figure}
\centering
\psfig{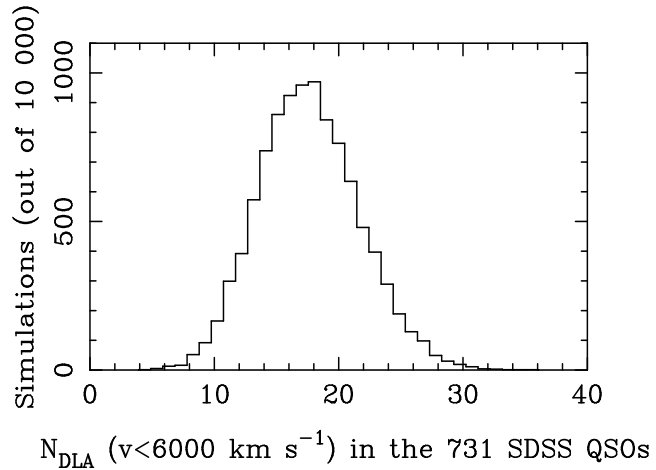}
\caption{\label{mc_test}Distribution of numbers of PDLAs found in Monte Carlo
simulations of samples of 731 QSOs with the same redshift distribution as the
SDSS sample.  PDLAs
have been included with the $n(z)$ determined by Storrie-Lombardi
\& Wolfe (2000).  Errors in $n(z)$ have been included based on the
Storrie-Lombardi \& Wolfe error bars drawn from a Gaussian
distribution.  The actual number of PDLAs found in the SDSS was 33;
this number (or higher) occurred in 0.05 \% of Monte Carlo realisations. }
\end{figure}

We produced 10,000 realisations of the 731 QSO sample; in Fig. 
\ref{mc_test} we show
a histogram of $N_{DLA}$ corresponding to the 731 QSOs in row 5 of Table 3.  
The mean number of DLAs at $\Delta v < 6000$ \kms\ in our simulations
is  $N_{DLA}=18.2 \pm 4.2$, compared with the observed value
of $N_{PDLA}= 33$, a difference of 3.5 $\sigma$.  From this we conclude
that there \textit{is} a statistically significant excess of
DLAs at small velocity separations.

In fact, 33 or more PDLAs were found in only 0.05 \% of our simulations.
The inferred significance of there being an overdensity of DLAs in 
each SDSS QSO sample is shown in the footnotes of Table 3.  
The redshifts of 
each QSO in the samples from the literature are unavailable and so the 
significance of the overdensity could not be calculated in these samples.

\section{Discussion}

\subsection{An Excess of PDLAs at $\Delta v < 6000$ \kms and dependence on
Radio-Loudness}

We begin by considering the number density in the full 731 SDSS
QSO sample used in this work to address the first question posed
in the introduction: is the density of DLAs in the vicinity of
QSOs significantly higher than that along the line of sight?
From rows 5 and 6 of Table 3 we find a factor of $\sim$ 2 overdensity
of $n(z)_{PDLA}$ compared with intervening systems
for velocities $<$ 3000 and $<$ 6000 \kms (overdense at 3.5$\sigma$
significance for $\Delta v <$ 6000 \kms).

Dividing our sample between radio-loud and radio-quiet QSOs we can
address the second question of the introduction: do the environments
of RQQs differ from those of RLQs?
For RQQs within $\Delta v <$ 6000 \kms\ (row 1 of Table 3), the measured
$n(z)_{PDLA}$ exceeds $n(z)_{DLA}$ (or $N_{PDLA}$ exceeds $N_{DLA}$)
by a factor 1.8; an overdensity of DLAs at the 3.0 $\sigma$ level. 
For $\Delta v <$ 3000 \kms\ (row 2 of Table 3), an excess is also apparent,
but only significant at the 1.4 $\sigma$ level.
A similar excess of PDLAs is found for RLQs ($\Delta v <$ 6000 \kms), 
but is significant at just the $\sim$1.4 $\sigma$ level, due to the small
number of
PDLAs in the sample (4 PDLAs; Table 3 row 3).  We find just one PDLA within
$\Delta v <$ 3000 \kms\ amongst the RLQs in our sample, but by combining our
result with those from the CORALS (Ellison et al. 2002) and FBQS (White et al.
2000) samples of RLQ PDLAs, we raise this number to five (row 4 of Table 3),
and find again that $n(z)$ exceeds the value for intervening absorbers, this
time by a factor 2.8 (no error quoted on the FBQS sample).

We therefore confirm the excess of DLAs close to the redshift of RLQs 
seen by Ellison et al. (2002), but we find that a similar excess is present 
in the vicinity of RQQs with no significant difference between the 
two environments. 

The excess of absorbers within 6000 \kms\ of both RQQs and RLQs
may be due to galaxies in the same cluster or supercluster as the QSO.
Overdensities of galaxies associated with quasars have been noted by
many authors (see \S1).
However, velocity distributions of galaxies within individual clusters 
at z$\sim$0 are typically
$\sim$ 1000 \kms, so the apparent overdensities found here
($\Delta v <$ 6000 \kms) imply structure on supercluster scales. 
Calculating the physical distance between QSOs and PDLAs is not
straightforward, particularly if both are located in a large
mass potential where peculiar velocities dominate.  
We can only calculate the line of sight separations
under the assumption that the velocities are driven by the
Hubble flow.  In this case (which of course is not valid
if the overdensities are linked to clustering) we find
QSO-PDLA separations up to 70 comoving Mpc.

Many authors have reported galaxy clustering around QSOs
on scales of a few comoving Mpc 
(e.g. S\"{o}chting, Clowes \& Campusano 2004 and references therein) 
whilst QSOs themselves
seem to cluster over similar distances (Croom et al. 2004;  
Porciani, Magliocchetti \& Norberg 2004).  Even larger
scale structures have been reported ranging from tens (Steidel et al. 1998; 
S\"{o}chting, Clowes \& Campusano 2002; Haines et al. 2003) to 
hundreds (Quashnock, Vanden Berk \& York 1996; Loh, 
Quashnock \& Stein 2001; Haines, Campusano
\& Clowes 2004) of comoving Mpc.  Large scale structures of many Mpc
are therefore pervasive from the lowest (e.g. Geller \& Huchra 1989) to
the highest (z $\sim$ 6; Stiavelli et al. 2005) redshifts observed.  
An excess of PDLAs, if the overdensities can be linked to clustering, is not
surprising given the considerable body of work that supports galaxy clustering
over such large volumes.  The extent of the overdensity, in the range
$\sim$ 40\% to a factor of two, is also consistent with what has been seen in
galaxy surveys (e.g. Sanchez \& Gonzalez-Serrano 1999;
Williger et al. 2002; Haines et al. 2004; Overzier et al. 2005).
Our observations of PDLAs could therefore confirm the scale and extent
of galaxy excesses that have been previously mapped in 2 dimensions.

\subsection{The Neutral Gas Content of PDLAs}

\begin{figure}
\centering
\psfig{file=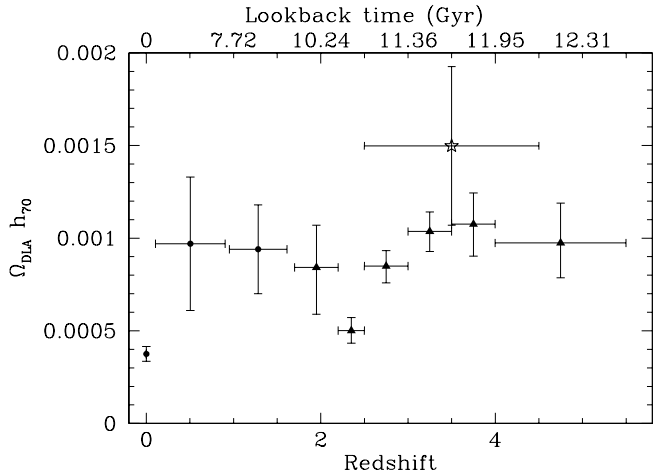,width=9.5cm,angle=270}
\caption{\label{omega} Mass density of neutral gas as a function of redshift.  Intervening
DLAs from the SDSS DR3 (Prochaska et al. 2005) are shown as filled
triangles.  Low redshift (Rao 2005) and $z=0$ (Zwaan et al. 2005) data
are shown with filled circles.  The value determined for PDLAs (this
work) in DR3 is shown with an open star.}
\end{figure}

In addition to the number density of PDLAs, we can also calculate
the mass density of neutral gas as a fraction of the closure
density, $\Omega_{DLA}$:

\begin{equation}
\Omega_{\rm DLA} = \frac{H_0 \mu m_H}{c \rho_{crit}} 
\frac{ \sum_{i} N_i({\rm H~I})}{ \Delta X}  
\end{equation}

where $\mu$ is the mean molecular weight (= 1.3) and $m_H$ is
the mass of a hydrogen atom.
Note that only neutral gas contributes to this quantity, so gas that
has been ionised, e.g. due to the proximity of the QSO is not included
in the sum.  For our adopted
cosmology $X$ and $z$ are related by the equation

\begin{equation}
X(z) = \int_0^z (1+z)^2[(1+z)^2(1 + z\Omega_M)-z(2+z)\Omega_{\Lambda}]^{-1/2}
\end{equation}

$\Delta X$ is therefore calculated analogously to $\delta z$
by summing over the 731 QSOs in our sample. We determined
log
$\Omega_{DLA} h_{70} = -2.82^{+0.110}_{-0.146}$, with
errors calculated
as in Storrie-Lombardi, Irwin \& McMahon (1996).
In Fig. \ref{omega} we compare the values determined for
$\Omega_{DLA}$ in the intervening DLA population in the DR3
(Prochaska et al. 2005) and at low redshift (Rao 2005;
Zwaan et al. 2005)
with the value determined for PDLAs. The PDLA value is
$\sim$50\% larger
than for the intervening systems, but also is consistent within
the large error
bars with the DLA value. If the higher $\Omega_{DLA}$ for PDLAs is confirmed,
this may simply reflect the
excess of the number density, $n(z)$, if the distribution of column
densities of
neutral hydrogen are similar. However,
since $\Omega_{DLA}$ is dominated by the rare high \nhi\
systems, the gas mass density is very sensitive to small
sample
sizes. We have 33 PDLAs in our sample, whereas each of the
$z>2$
SDSS bins typically contains 100 -- 200 DLAs. If we
exclude our highest
column density system from the calculation, $\Omega_{DLA}$
for PDLAs is in excellent agreement with the intervening
systems. Similarly, if
we include a further high-\nhi\ system, $\Omega_{DLA}$ for
PDLAs is $\sim$
twice that of the intervening DLAs, a result consistent
with the observed
excess of proximate DLAs and similar sample average \nhi.

\subsection{The Systemic Redshift of QSOs and PDLA Velocity Distribution}

Our definition of PDLAs is based on relatively small velocity offsets,
which in turn are dependent upon accurate systemic redshifts.
It has been known for decades that there are systemic shifts between 
the high ionisation broad emission lines (e.g. Gaskell 1982;
with recent reviews by Richards et al. 2002 and Vanden Berk et al 2001).  
The systemic redshifts of QSOs in SDSS are 
determined from a cross-correlation of the spectrum with the composite 
of Vanden Berk et al. (2001), which itself is constructed by assuming 
the forbidden [O~III] line (5007\AA) is representative of
the true systemic redshift.  This technique minimises the systemic 
offset that may occur when measuring the redshift of a QSO from the 
\lya\ emission line if the line is blended by a PDLA.  However, the 
scatter in blue/redshifts of \lya\ (in particular, since it is the 
strongest emission line covered in our redshift
range) relative to [O~III] will contribute an uncertainty in 
the redshift.  This scatter between $z_{\rm Ly\alpha}$ and 
$z_{\rm [OIII]}$ is relatively small however; from 
Vanden Berk et al. (2001) $\Delta v = 143 \pm 91$ \kms.

\begin{table}
\centering
\caption{$n(z)_{PDLA}$ derived for different velocity ranges, to check the
effect of systematic errors in velocity.}
\begin{tabular}{ccc}
\hline
$\Delta v$ (\kms)&$N_{PDLA}$&$n(z)_{PDLA}$\\
\hline
0 - 3000   &12&0.38$^{+0.14}_{-0.11}$\\ \rule[0mm]{0mm}{4mm}
100 - 3100 &12&0.38$^{+0.14}_{-0.11}$\\ \rule[0mm]{0mm}{4mm}
200 - 3200 &16&0.51$^{+0.16}_{-0.13}$\\ \rule[0mm]{0mm}{4mm}
500 - 3500 &17&0.54$^{+0.17}_{-0.13}$\\
\hline
\end{tabular}
\normalsize

The errors associated with $n(z)_{PDLA}$ are 1$\sigma$ confidence limits
(Gehrels 1986).
\end{table}

In order to assess the impact of uncertainties
in $z_{\rm sys}$ we have calculated $n(z)_{PDLA}$ in different velocity bins
where the lower limit (always assumed to be zero in our original 
calculations) is
offset by 100, 200 and 500 \kms\ ($\Delta v$ is held constant).  
The results are presented in Table 5.  
Although there is some variation in the different velocity bins, the
number densities are consistent with each other within the Poisson errors 
(again, calculated from Gehrels 1986).  Therefore, although an individual
QSO may have an error of several hundred \kms\ in its quoted systemic 
redshift, this is unlikely to reproduce an excess of PDLAs within several
\textit{thousand} \kms.

Fig. \ref{v_hist} 
shows a histogram of the distribution of PDLA velocities relative 
to $z_{sys}$. We have binned the histogram relatively coarsely ($\Delta$
1000 \kms) due to the errors in redshift discussed above and also
the modest sample size.  The  apparent peak of the distribution lies 
at $\sim$3500 \kms, and the intervening DLA number density is recovered 
by $\Delta z\sim$ 6000 \kms.  However, the position of the peak is not 
highly significant, and the distribution suffers from low number 
statistics.  Despite the 
low number statistics it is interesting to
note that Williger et al. (2002) also found \textit{transverse}
overdensities between MgII absorbers and QSOs on scales 3000--4500 \kms\
(albeit in one velocity direction only).

Prochaska et al. (2005) published a statistical sample of
SDSS DLAs in the DR3 which excludes proximate systems.
However, this cut was done \textit{a posteriori} and PDLAs
were identified in their `full' DLA list. Although the
selection criteria of the Prochaska et al. sample are
somewhat different to ours (e.g. based on colour selection,
S:N redshift range and FIRST coverage), we can calculate
relative numbers of PDLAs to intervening DLAs from their
full sample (Prochaska, 2005, private communication) which is
continuous in velocity distribution. The overdensity of
PDLAs found by us is replicated in the Prochaska et al. sample which
contains 50 DLAs at $\Delta$v$<$3000 \kms\ \footnote{This
number is larger than the number of PDLAs in our sample because of 
the less stringent selection conditions imposed by Prochaska et al.
(2005) on their parent QSO sample.}, a factor of $\sim$ 10 more than
found in similar velocity width bins at $v >> 6000$ km/s. The
Prochaska et al. sample also confirms that by v=6000 \kms\ the
number density has fallen to the background level.

In theory, the distribution of galaxy velocities in the
field of a given QSO can provide information of the environment, such
as the mass potential.  Here, however, the interpretation of velocity
distribution is not straightforward. Each PDLA is likely to inhabit
a different environment and the QSOs possess a range 
of properties such as luminosity and radio-loudness.  
With a large number of absorbers, distinct populations may be discerned.
For example, Weymann et al. (1979) produced
a histogram like Fig. \ref{v_hist}, for a large number of C IV absorbers
and divided the populations into `intervening' and `ejected'
components.  Larger samples of PDLAs may be used to distinguish
the velocity distributions of absorbers associated with galaxy
overdensities near QSOs.  With the final releases of SDSS data to follow, 
a further 2400 degrees$^2$ of spectroscopic sky coverage (and hence many 
PDLAs) is expected to be available.  It will then be possible to measure
the excesses we find more accurately, reducing the errors in $n(z)$. 
It may also
be possible to measure $n(z)$ as a function of
velocity from the quasar.  With the current statistics we can only
note that by $v \sim 6000$ \kms\ the number density is close to that
of the intervening DLAs.

Finally, we note that the clustering of DLAs around QSOs could be
independently checked by observing QSO pairs with slightly different
redshifts, but modest impact parameters (e.g. Hennawi et al. 2005).  
If DLAs preferentially occur near QSOs, we would expect 
to see the same overdensity of DLAs at approximately the redshift of the nearer
of the pair imprinted on the spectrum of the more distant QSO.

\section{Conclusions}

We have searched the spectra of 731 SDSS DR3 2.5 $< z_{\rm sys} <$ 4.5 QSOs 
for DLAs (\nhi\ $> 2 \times 10^{20}$ \cm2) and found 33 within 
$\Delta v <$ 6000 \kms\ (and 12 within $\Delta v <$ 3000 \kms).
Our search increases the number of PDLAs reported in the literature 
within $\Delta  v <$ 3000 \kms\ (the conventional definition of
$z_{\rm abs} \sim z_{\rm sys}$ DLAs) from 5 to 16.

\begin{figure}
\centering
\psfig{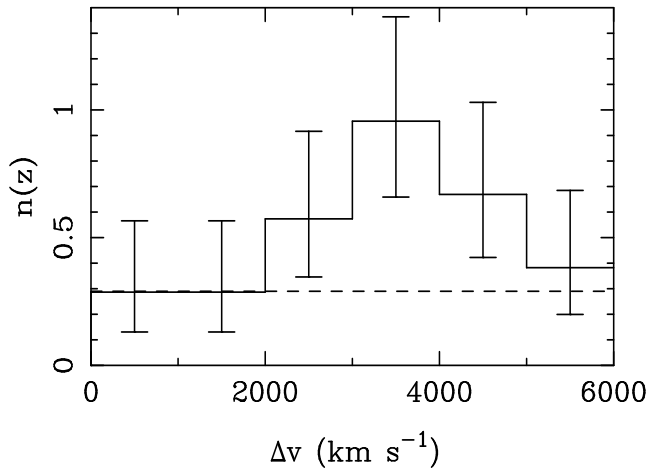}
\caption{\label{v_hist} $n(z)_{DLA}$ found in SDSS as a
function of $\Delta v$.  The bins are 1000 \kms\ wide.  The dashed line 
indicates the expected number density for intervening DLAs $n(z)_{DLA}$.}
\end{figure}

The number density of DLAs within $\Delta v <$ 6000 \kms\ of the 
redshift of the QSOs is approximately double that observed for samples 
of intervening DLAs (a) found in
the same sample, the SDSS (Prochaska et al. 2005) and (b) for the 
relation derived by Storrie-Lombardi \& Wolfe (2000).  The excess is
significant at the 3.5 $\sigma$ level.
There is no significant difference between the number density of
PDLAs towards RLQs and RQQs.
A similar excess of PDLAs is found within $\Delta v <$ 3000 \kms\ of QSOs;
$n(z)_{PDLA}$ / $n(z)_{DLA}$ = 1.7 for all RQQs and RLQs combined.  We
speculate that  the overdensities of DLAs in the vicinity 
of QSOs may trace galaxies in the same
clusters or superclusters as the QSOs. The similar excesses around radio-quiet
and radio-loud quasars are consistent with their inhabiting similar
environments, as suggested by recent observations (e.g. Finn, Impey \& Hooper
2001; Wold et al. 2001).  This emphasises the importance of imposing a 
velocity cut when studying intervening DLAs, not only to avoid intrinsic 
AGN absorption, and any proximity effect, but also to avoid objects 
in the cluster or supercluster hosting the QSO.  Moreover, we
recommend that, based on Fig. \ref{v_hist}, $\Delta v$
$\sim$ 6000 \kms\ is a suitable velocity cut.  

A number of PDLAs in the DR3 exhibit strong metal lines, despite
the moderate resolution and S:N ratios of the spectra.
With high-resolution spectroscopy, it should be possible
to determine whether the metallicities of PDLAs are enhanced 
compared with intervening DLAs, as may be expected if they are 
associated with the QSO environment.  For example, Ellison \& Lopez
(2001) and Lopez \& Ellison (2003) found unusual relative abundances
in a pair and a triplet of DLAs whose velocity separations were
$<10$ $000$ \kms.  Many QSOs in the SDSS are bright enough to follow-up
with an echelle spectrograph on an 8-m telescope, such as UVES on the
VLT.  Investigating the abundances of PDLAs will be an interesting,
and so far unexplored, extension of this work.

\vspace{5mm}
$Acknowledgements$.
We are grateful to Jason X. Prochaska for providing the statistics 
for the intervening DLAs discovered in Data Release 3 prior to publication, and to
Patrick B. Hall for valuable advice on the SDSS database.
Funding for the creation and distribution of the SDSS Archive has been provided by
the Alfred P. Sloan Foundation, the Participating Institutions, the National
Aeronautics and Space Administration, the National Science Foundation, the U.S.
Department of Energy, the Japanese Monbukagakusho, and the Max Planck Society. The
SDSS Web site is http://www.sdss.org/.

The SDSS is managed by the Astrophysical Research Consortium (ARC) for the
Participating Institutions. The Participating Institutions are The University of
Chicago, Fermilab, the Institute for Advanced Study, the Japan Participation Group,
The Johns Hopkins University, the Korean Scientist Group, Los Alamos National
Laboratory, the Max-Planck-Institute for Astronomy (MPIA), the Max-Planck-Institute
for Astrophysics (MPA), New Mexico State University, University of Pittsburgh,
University of Portsmouth, Princeton University, the United States Naval Observatory,
and the University of Washington.

\end{document}